\documentclass[usenatbib]{article}
\usepackage{times,graphicx,natbib,a4wide}

\begin{document}
	
\title{Extrasolar Asteroid Mining as Forensic Evidence for Extraterrestrial Intelligence}
\author{Duncan H. Forgan$^1$ and Martin Elvis$^2$}
\maketitle

\noindent $^1$Scottish Universities Physics Alliance (SUPA), Institute for Astronomy, University of Edinburgh, Blackford Hill, Edinburgh, EH9 3HJ, UK \\
\noindent $^2$Harvard Smithsonian Center for Astrophysics, 60 Garden Street, MS 6, Cambridge, MA 02138, USA

\noindent \textbf{Word Count: 5,070} \\

\noindent \textbf{Direct Correspondence to:} \\
D.H. Forgan \\ \\
\textbf{Email:} dhf@roe.ac.uk \\
\textbf{Post:} Mr Duncan Forgan \\
East Tower, Institute for Astronomy \\
University of Edinburgh \\
Blackford Hill \\
EH9 3HJ, UK

\newpage

\begin{abstract}

\noindent The development of civilisations like ours into spacefaring, multi-planet entities requires significant raw materials to construct vehicles and habitats.  Interplanetary debris, including asteroids and comets, may provide such a source of raw materials.  In this article we present the hypothesis that extraterrestrial intelligences (ETIs) engaged in asteroid mining may be detectable from Earth.  Considering the detected disc of debris around Vega as a template, we explore the observational signatures of targeted asteroid mining (TAM), such as unexplained deficits in chemical species, changes in the size distribution of debris and other thermal signatures which may be detectable in the spectral energy distribution (SED) of a debris disc.  We find that individual observational signatures of asteroid mining can be explained by natural phenomena, and as such they cannot provide conclusive detections of ETIs.  But, it may be the case that several signatures appearing in the same system will prove harder to model without extraterrestrial involvement.  Therefore signatures of TAM are not detections of ETI in their own right, but as part of ``piggy-back'' studies carried out in tandem with conventional debris disc research, they could provide a means of identifying unusual candidate systems for further study using other SETI techniques.

\end{abstract}

\section{Introduction}

\noindent The search for extraterrestrial intelligence (SETI) has been primarily concerned with detecting artificial radio signals as a means of confirming the presence of extraterrestrial intelligences (ETIs).  While a necessary and obvious search method, the science of SETI can only benefit from developing a multi-wavelength, multi-signal approach, such as optical SETI \citep{Mead2010,Werthimer2010}, the search for extraterrestrial artifacts such as Dyson spheres \citep{Dyson}, and the more mainstream searches for potentially habitable planets such as the \emph{Kepler} \citep{Kepler2010} and \emph{Mearth} missions \citep{Nutzman2009}.  

Artificial signals derived from well-studied astrophysical objects are particularly desirable, as they provide fertile ground for so-called ``piggy-back'' searches, which are in general easier to justify than dedicated searches which have more limited secondary science goals.  One class of astrophysical object currently enjoying significant attention is the \emph{debris discs} surrounding evolved stars.  The remnants of more massive, gaseous discs that encircle young stars during their formation and initial evolution, they are composed of rocky/icy debris in a distribution of sizes.  Like the belts of asteroids, comets and other bodies found in our own Solar System, such debris may be the ``leftovers'' from planet formation, and is expected to be common in planetary systems, with lifetimes of order tens of millions of years after the star's formation.  

Debris discs are typically detected in the infrared (IR) and sub-millimetre regimes, using photometry, spectroscopy or imaging (\citealt{Wyatt2008,Krivov2010} and references within).  Therefore, they have been ideal candidates for study using space-based and ground-based telescopes over the past 30 or so years, beginning with the first detection of a debris disc around Vega \citep{Aumann1984}.  They can be used as forensic evidence of earlier planet formation, and they may even confirm the presence of planets due to dynamical features such as clumping and resonances (e.g. \citealt{Wyatt2002,Greaves2005,Greaves_Tau}).  With new data arriving from the recently commissioned \emph{Herschel} Space Telescope (e.g. \citealt{Vandenbussche2010}), and the wealth of data generated from its predecessor, \emph{Spitzer}, it would be advantageous for SETI researchers if debris discs could provide artificial signals indicating extraterrestrial intelligence.

Targeted asteroid mining (TAM) could provide such evidence.  Engineering limitations experienced by mankind will be the same for ETIs with a similar evolutionary history - considering the large quantities of raw material required to build space vehicles and habitats, TAM may be a difficult, but necessary (or at least highly desirable) skill for civilisations to advance along the path to becoming a truly spacefaring civilisation.   This would appear to be true for both biological and post-biological civilisations \citep{Cirkovic_postbio2006,Dick2008}, as both require construction and manufacturing as a means of sustaining themselves.

We propose that the deliberate extraction of specific raw material by extraterrestrial intelligences (ETIs) from a debris disc may provide a variety of artificial observational signatures, detectable in the infrared (amongst other wavelengths), similar to the ``Dysonian'' signatures suggested by \citet{Cirkovic_macro} for macro-engineering projects.  The aim of this paper is to investigate the nature of these signatures.  We will consider the effects of targeted asteroid mining on the properties of the Vega debris disc system (while simultaneously considering putative future attempts by humans to mine in the Solar System).  The paper is laid out as follows:  in section \ref{sec:mining} we consider asteroid mining and its potential as a technology driver for space exploration.  In section \ref{sec:debrisdisc} we introduce the debris discs, focusing particularly on the debris disc surrounding Vega.  We discuss the potential observational effects of asteroid mining in section \ref{sec:features}.  We finally consider how likely it would be for these signals to be detected and successfully characterised as artificial using current instrumentation.

\section{The Case for Prevalent Asteroid Mining \label{sec:mining}}

\noindent Planets have finite natural resources.  This truism has become painfully apparent to mankind in recent decades, through examples such as shrinking biodiversity and the increasing challenges facing engineers and geoscientists attempting to extract fossil fuels from the Earth.  All life acts as consumers at some level, but the level of consumption is typically regulated through population control and other pressures introduced by the ecosystem .  Advances in technology have allowed humans to circumvent these controls, with the effect that humans have vastly increased their population, placing strains on local resources.  There has also been a continued increase per capita in consumption of precious metals for technologies such as computers, mobile phones and the infrastructures which enable them to function.  The proposed green technologies of the future, such as hydrogen fuel cells and CO$_2$ scrubbers, will only enhance this need for already rare resources \citep{Elshkaki2006,Schuiling2006}.

Such resources can be found in the asteroids.  Meteoritic analysis \citep{Kargel1994}, suggests that large quantities of gold, platinum and other precious metals exist in the asteroids of the Solar System, as well as large amounts of other elements such as iron, nickel, magnesium and silicon. He concludes that successful operations at modest mining rates could increase the total production rate of some materials by a factor of 10.  By applying simple empirical models (where market value scales as the square root of production rate), approximate threefold decreases in price can also be expected, over timescales of a few decades.  Indeed, if the supply of precious metals such as platinum is to continue to meet technological demands, asteroid mining may become essential within the coming century \citep{Elshkaki2006}.  Besides these industrially driven arguments, SETI scientists are driven by the possibility of detecting extraterrestrial intelligence by evidence of their activities in the Outer Solar System and the asteroids \citep{Papagiannis1978,Papagiannis1995}.  Developing asteroid mining technology for commercial reasons will certainly assist the implementation of studies of this nature.

Humans have not begun asteroid mining primarily for reasons of political economy.  While the resources still exist in affordable quantities on Earth, governments lack a good short-term economic case to attempt dangerous missions at high cost to bring back what would initially be modest quantities of raw materials.   As  \citet{Hickman_economy} observes, asteroid miners should not expect immediate investment from private investors either.  While the potential return from successful, properly matured asteroid mining missions is very large, the level of capital required up front for any large-scale space project is also very large - \citet{Schmitt1997} optimistically estimates a sum of around \$15bn for general commercial space enterprise (assuming fusion technologies based on lunar $^3$He become profitable, and not considering the problems presented by the current financial landscape).  Further to this, the maturation time period (before profits can be generated) is too long, i.e. greater than 5 years.  Other large-scale space projects (such as Martian colonisation) are equally unappealing for investors looking for returns on their investments within a decade - \citet{Hickman_economy} gives a simplified example which shows that if Mars can be terraformed in less than a thousand years, even a modest rate of interest on an initial loan requires Martian real estate to be extremely expensive.

While there might not be a good short-term economic case for governments to fund TAM missions, there are long-term economic and political motivations (see \citealt{Gerlach2005} for a thorough review).  If the initial high capital barrier can be overcome, and profits can be generated, then manufacturing future technologies will become much cheaper as the precious metals become less precious.  The expertise gained by designing and undertaking TAM missions can then be brought to bear on other challenges in space exploration.  Given the hazards involved in TAM for atmosphere-breathing species, it is reasonable to assume that much of the process will become automated and autonomous, ushering in a new era of robotics with advanced decision-making and goal-seeking software (which has obvious implications for post-biological evolution).  With a large surplus of raw materials and a skilled robotic workforce, large, permanent space habitats can be constructed, for example in geostationary orbit.  This may allow the construction of the long-considered ``space elevator'' (cf \citealt{Aravind2007}), greatly reducing the cost of space-travel in general.  With much cheaper space exploration, the financial risks are reduced for other large-scale space projects, facilitating capital investment and Man's continued development into a space-faring species.

Governments which invest at early stages in these projects will receive profitable advantages over their competitors, including early access to raw materials, new technologies and highly skilled personnel, each a boost to any nation's economy.  These benefits may not outweigh the current financial disadvantages, but dwindling resources and rising costs on Earth will gradually improve the prospect of developing TAM missions until they become an obvious choice.  Exactly how the initial capital will be raised will be the most important and difficult obstacle - private investors will baulk at the prospect of entirely funding TAM, but as with other large scale projects such as the Panama Canal (which was also faced with technological challenges and capital problems) the action of governments can make all the difference, especially if they can be encouraged into competition with each other \citep{Hickman_economy}.

While this argument is clearly Earth-oriented, much of it applies in general to intelligent species which consume planetary resources at sufficient rates.  ETIs which have similar economic concerns to ours will eventually find extraplanetary mining projects desirable as their own resources become depleted (provided of course they are sufficiently technologically advanced).  We suggest the complexity of TAM missions are such that most species capable of it have the potential to become truly space-faring.  If technological civilisations more advanced than ours exist in the Galaxy, a distinct possibility given the estimated median age of terrestrial planets being around 1 Gyr older than Earth \citep{Line_planets}, and asteroid mining is a common activity which underpins their existence, then searching for signatures of TAM is an appropriate activity for SETI to undertake.  However, we must balance this with the realisation that systems 1 Gyr older than Earth will most likely no longer have debris discs.  Exactly why this is the case will be explained in more detail in the next section.

\section{Debris Discs \label{sec:debrisdisc}}

\noindent Debris discs are composed of solid material, in bodies hundreds of kilometres in size (i.e. dwarf planets) to sub-micron sizes (i.e. dust).  These objects orbit the central star, initially in a narrow belt often referred to as the ``birth ring'', typically at distances of tens to hundreds of AU (Astronomical Units, where 1 AU = semi-major axis of the Earth's orbit).  They are primarily subject to the forces of gravity (which acts radially inward towards the star and giant planets), and radiation pressure (which acts radially outward from the star).  The efficiency of radiation pressure typically depends on the size of the object, with dust grains feeling a relatively stronger outward force than the larger bodies, generating more eccentric orbits.  For sufficiently small grains, the radiation pressure will defeat gravity, and \emph{blow out} the grain from the system.  Stellar radiation also introduces tangential forces on the solids, giving rise to Poynting-Robertson drag, which removes angular momentum, and the Yarkovsky effect (which will add angular momentum provided the dust's orbit is prograde). Typically, Poynting-Robertson drag will dominate, resulting in inward radial motion and allowing dust to be deposited on the star itself (see \citealt{Krivov2010} for a review).

However, debris discs still possess large amounts of dust, which comes into thermal equilibrium with the local radiation field, absorbing and re-emitting photons at infrared wavelengths.  As dust is continually removed from a debris disc system by blow-out and drag, a source of dust production must be identified.  Collisions are typically invoked to provide such a source, requiring sufficiently large relative velocities amongst the debris.  Debris discs form from the remains of more gaseous protoplanetary discs, and hence gas drag will initially damp any relative velocities in the debris population. The debris must therefore be ``stirred'' by some mechanism, e.g. by nearby planets, or ``self-stirred'' by larger debris, for collisions to be initiated \citep{Wyatt2008}. Once this is achieved, a collisional cascade begins.  Larger bodies will collide at a variety of different impact angles and size ratios, leading to a variety of possible outcomes, from erosion to cratering to complete disruption.  This process grinds down larger bodies through intermediate sizes down to dust grains, after which blow-out and drag can remove them from a system.

A simplified quasi-steady state for the debris disc can be constructed assuming that collisional cascade produces sufficient dust to balance dust losses due to blow out and drag.  This generates a constant probability distribution of grain size $a$ \citep{Muller2010}:

\begin{equation} P_{coll}(a) \propto a^{-3.5} \end{equation}

\noindent These assumptions hold in the limit where the collision timescale is short compared to the characteristic timescale of the drag forces involved, which is usually the case for debris discs (and is true for Vega, the system we will study).  The state is quasi-steady as the mass of the system must decrease as the destroyed larger bodies are not replenished.  This steady state can be perturbed by singular events such as the occasional break up of particularly large objects, or instabilities generated from nearby planets, but evidence of these perturbations will disappear as equilibrium is re-established. A debris disc's ability to ``refresh'' after such perturbations weakens with age, therefore we may expect older systems to exhibit these perturbations more readily.

The radial distribution of grains varies with the grain size.  Grains of a given size will evolve away from the birth ring, with some remaining in the birth ring while some are blown by radiation pressure into eccentric orbits (and larger apastron radii).  Smaller grains are more susceptible to radiation pressure, and hence move on more eccentric orbits.  The equilibrium distribution shows grains close to the blowout size occupying the widest range of radii, with the radial range decreasing with increasing grain size (see \citealt{Muller2010} for details).

A related structure in a planetary system is the \emph{exozodiacal dust} (or exozodi for short).  Often considered as an extension of the debris disc into smaller orbital radii, they are the analogues of zodiacal dust in the Solar System (zodi), produced again by collisions.  This dust (of sizes $\sim 1 - 100 \mu m$) settles into resonant ring configurations around the inner planets, with densities of order seven times above background values \citep{Stark2008} with optical depths $\sim 100-1000$ times larger than the solar zodi.  As they tend to occupy resonances close to potentially habitable planets, they may also be fertile ground for mineral extraction.

\section{Observational Features of Asteroid Mining \label{sec:features}}

	\subsection{Chemical Disequilibrium}

\noindent The ultimate purpose of TAM is to extract as large a quantity of specific chemical species as is possible while expending the least amount of energy possible.  In much the same way that Earth reveals its biosphere through an atmosphere in apparent chemical disequilibrium \citep{Lovelock1975,Catling2010}, it is intuitively obvious that asteroid mining will induce an imbalance in the abundances of species in a debris disc (relative to expected values, which may be ascertained by considering the metallicity of the parent star and comparing with solar abundances).

As we cannot directly assay extrasolar debris as we do with meteorites on Earth, we must restrict ourselves to observations based on the light we receive from them.  The debris belts in our Solar System exhibit signatures of many useful substances, some of which have been detected in extrasolar debris discs.  Silicates are commonly detected (e.g. \citealt{Schutz2005}), which are used by mankind for ceramics, solar panels and other computer technology.  Indeed, olivine (a magnesium-iron silicate thought to be found in the mantle of Vesta, one of the larger asteroids, \citealt{Shestopalov2008}) is of value to efforts to sequester CO$_2$ \citep{Schuiling2006}.  Amorphous silicates generate a characteristic emission peak at 10 $\mu m$, due to UV starlight being absorbed on the disc surface by silicate grains and re-emitted.  It is tempting to ascribe an absence of this silicate feature to TAM, but such an inference cannot be justified, as thermal annealing can reduce the available amorphous silicates by converting them into crystalline forms such as forsterite \citep{Hallenbeck1998}.  Tentative detections of extrasolar water ice, carbonates and polycyclic aromatic hydrocarbons (PAHs) are also reported \citep{Chen2008}, whose uses are obvious to mankind.  

However, silicates, carbonates and water are found in large quantities on Earth.  The same is expected to be true for most terrestrial planets, and therefore the socio-economic pressure to specifically mine these substances will typically be low.  Elements which differentiate efficiently (i.e. the heavier elements such as iron and nickel) are more probable targets, as differentiation will in general remove them from easy reach on terrestrial planets.  They are of more practical use in large-scale space engineering projects, where large masses of material are required.  By comparison, building solar panels with silicates requires relatively low quantities of material, even for large-area projects.  Rarer elements such as platinum, palladium, indium and germanium are also appropriate targets for their use in technological innovation (and their expected rarity in terrestrial planets).

Detections of the more desirable heavy elements prove difficult, as modelling spectral energy distributions (SEDs) in order to inventory chemical species is degenerate, i.e. any one spectra can be explained by a variety of different mixtures.  If the Solar System's contents are any indication, then debris discs will hold a rich variety of species in relatively large quantities, including nickel, iron, platinum and many other useful elements \citep{Kargel1994}, however current observations do not in general detect these materials conclusively due to the aforementioned degeneracies.  High resolution spectroscopy will be required if any detection is to be made.  Iron and nickel produce spectral lines in the near ultraviolet - the young disc system $\beta$ Pictoris has yielded transient detections of these species and other metals in absorption.  The velocities of the lines have been successfully modelled by the infall of evaporating comets (see \citealt{Madjar1998,Karmann2001}).  Detection of such lines with unusual velocity profiles (or an unexplained absence of them) would be another indicator of potential artificial activity.

While absences of any one of these chemical species can in general be explained by natural phenomena, the depletion of several of these species becomes harder to explain (although not impossible).  While not conclusive proof of TAM, if such characteristic deficits - in what are elements of potential utility by ETIs - cannot be satisfactorily explained by other models, then TAM must be considered as a possibility.	

	\subsection{Mechanical Disequilibrium}

\noindent Mining is a destructive, energetic process - can this be detected in the dynamics of a debris-disc system?  The most obvious place to begin our search is the size distribution of objects.  It would seem sensible that TAM missions will preferentially select larger bodies over smaller bodies, both for the increased surface gravity to reduce operational complexity, and for the increased yields achievable in a single mission.  Persistent mining over long times will artificially reduce the number of larger bodies in the system, creating large numbers of small particles and dust due to mining inefficiencies as well as non-valuable debris (``tailings").  While large bodies are not directly detected by their IR flux, indirect inferences can be made by studying the dust.  

There are two important factors to determine if we are to make any useful predictions.  The first is the timescale on which TAM occurs, and the second is the size distribution of debris left by TAM.  Let us deal with these factors individually.

\begin{figure}
\begin{center}
\includegraphics[scale=0.7]{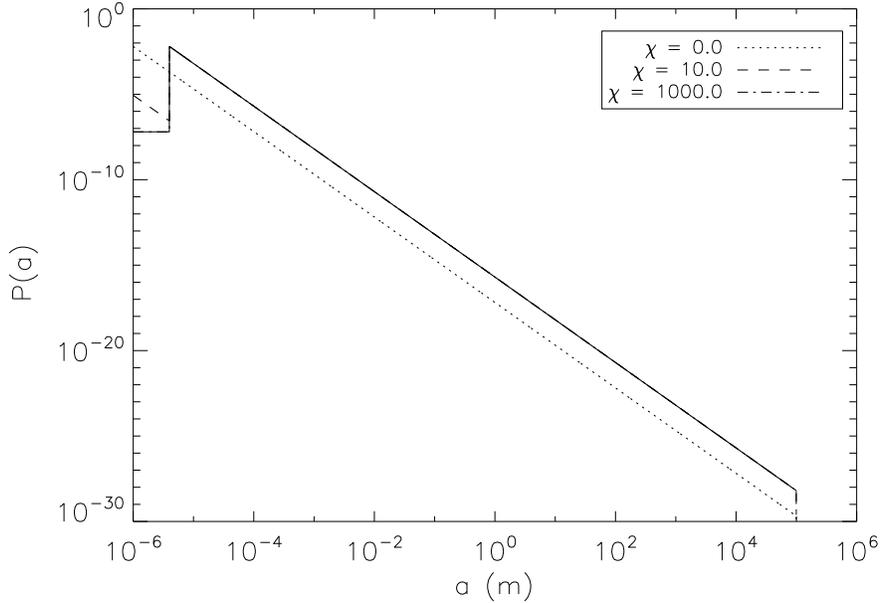}
\caption{The effect of asteroid mining timescale on the resulting debris population.  The distribution of mining debris $P_{mine}$ is equal to the distribution of the natural collisional debris $P_{coll}$.  The dotted line shows the effect of instantaneous mining ($\chi=t_{mine}/t_{blow}=0$), the dashed line a more moderate mining timescale ($\chi=10$), and the dot-dashed line a longer mining timescale ($\chi=1000$).  The solid line shows the initial debris population distribution before mining. Note that for $\chi=1000$, the initial and final distributions are indistinguishable.  $\chi$ is defined in the text. \label{fig:chi}}
\end{center}
\end{figure}

If we are to disrupt the steady state generated by the balance between collisions and dust losses - so that it may be detected on long-time scales, even when mining has ceased - then TAM must operate continuously on a sufficiently rapid timescale for perturbations to be long-lasting.  Figure \ref{fig:chi} demonstrates this concept - for simplicity we assume that the distribution of TAM debris $P_{mine}$ has the same power law dependence of collisional debris $P_{coll}$.  The solid line shows the steady state size distribution (for which we adopt the steady state distribution for Vega determined by \citealt{Muller2010}), and the dashed lines show the resulting distribution if the largest bodies are mined, and their debris is distributed according to a $a^{-3.5}$ power law.  To model the mining timescale, we define

\begin{equation} \chi = \frac{t_{mine}}{t_{blow}} \end{equation}

\noindent where $t_{mine}$ and $t_{blow}$ are the mining and blow-out timescales respectively.  Blow-out affects grains below $a\sim 4\,\mu m$ - we model blow-out during mining by reducing the proportion of grains in this critical regime by a factor $e^{-\chi}$.

Figure \ref{fig:chi} shows the resulting distributions for $\chi=$ 0 (instantaneous mining), 10 and 1000 respectively.  Instantaneous mining produces a large amount of dust, which would provide a detectable signature. As $\chi$ increases, this signature disappears until the steady state is essentially restored at $\chi=1000$ (although with depleted larger bodies).  The real value of $\chi$ depends on the scale of the mining mission.  Typical blow-out timescales are of order the disc crossing timescale, i.e. of order $10^2 - 10^3$ yr.  Let us assume mining of iron occurs at the current terrestrial rate of $10^9$ tons per year (see \citealt{USGS2010}).  To mine one 100 km asteroid at this rate (assuming 10\% of the mass is iron ore) would require timescales of $\sim 10^5$ yr.  This gives $\chi\sim 500-5000$, which would suggest that mining would not act quickly enough to disrupt the steady state.  However, we are also interested in future mining rates.  In particular, what might the mining rate be when Earth's iron reserves are depleted, which given current projections \citep{USGS2010} could be around a century from now? Global iron mining rates grew by approximately 3\% in 2009 - if such growth continues, it may be more appropriate to use a value of $\chi$ that is 10-100 times lower (which is certainly be possible if global consumption per capita grew to be on a par with that of the developed world).  This may put $\chi$ much closer to unity, and therefore dust production may be noticeable in the size distribution below the blow-out radius.

\begin{figure}
\begin{center}
\includegraphics[scale=0.7]{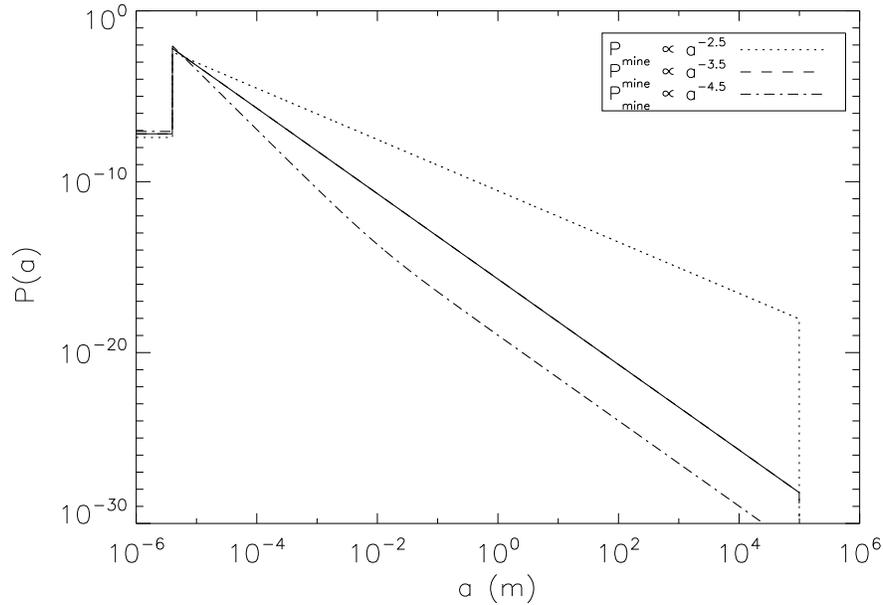} 
\caption{The effect of mining debris distribution. The mining timescale $\chi=500$ for all cases.  The dotted line shows $P_{mine}\propto a^{-2.5}$, the dashed line $P_{mine}\propto a^{-3.5}$, and the dot-dashed line $P_{mine}\propto a^{-4.5}$.  The solid line shows the initial debris population distribution before mining.  Note again that when $P_{mine}=P_{coll}$ and $\chi$ is large, the initial and final distributions of $\chi$ are indistinguishable. \label{fig:m}}
\end{center}
\end{figure}

But, we have not fully considered the distribution of debris produced by TAM, $P_{mine}$.  Ideally, there should be no debris, but mining is imperfect, and it is likely that material of limited use will be left in orbits similar to that of the target asteroid.  Figure \ref{fig:m} shows the effect that changing the power-law distribution of the mining debris has on the resulting size distribution (where we now take $\chi=500$).  

\noindent Even at the relatively slow mining rates currently achievable on Earth without ubiquitous robotic assistance, the effects of changing mining debris distribution becomes clear.  The dotted line in Figure \ref{fig:m}, shows the the effect of a mining strategy that tends to favour larger fragments more than the natural distribution ($P_{mine} \propto a^{-2.5}$).  These calculations conserve the volume of the asteroids being mined, and hence a significant effect can be found after the probability distribution is normalised.  However, this is an ineffective mining strategy - breaking up the debris into such large fragments may make for a dynamically unstable zone around the location of the asteroid, presumably resulting in uncontrolled collisions which could place the mining operation in jeopardy.  Also, throwing off such large fragments increases the probability that valuable resources are thrown away with it.

The dashed line represents the case discussed previously, where the mining debris and collisional debris distributions are equal.  As before, the large $\chi$ means that no artificial signatures may be seen.  The dot-dashed line represents a mining strategy that favours smaller fragments more than the natural distribution ($P_{mine} \propto a^{-4.5}$).  We can see that this affects the resulting distribution below $a\sim 1 \, cm$, steepening the power law.  This effect increases with a steepening power-law for $P_{mine}$.  Steep power law distributions have been detected, for example in HD172555, but they are well-explained by a hypervelocity impact between two bodies in the system \citep{Lisse2009}.

While favouring smaller fragments may present a better strategy based on dynamical arguments, the production of large amounts of dust is also hazardous to mining operations, clogging machinery and potentially causing failures.  For this reason, and for profit optimisation, terrestrial mining typically collects these ``tailings'' (i.e. $a < 1 \, mm$).   Such tailings could prove to be more hazardous in microgravity environments, and could cause a collisional avalanche, where dust production triggers further dust production as radiation pressure blows out the grains \citep{Artymowicz1997,Grigorieva2007}.  It would therefore seem imperative that tailings collection be efficient.
	
Considering these various factors, the true distribution of mining debris is unclear.  A single power-law description is presumably too simple - a two power-law description may be more appropriate, with a significant change around mm to cm sizes (i.e. the tailing size).  Whatever the true $P_{mine}$, its signature will be detectable in the size distribution (at least until the collisional cascade begins to reshape the distribution).  However, some signatures generated by TAM are easily mimicked by natural events, such as collisions betweeen very large bodies \citep{Krivov2010}.  Anomalous power-law distributions alone would therefore appear to be necessary but insufficient proof of TAM.

	\subsection{Thermal Disequilibrium}

\noindent If dust is produced as a result of mining, then the radial distribution of that dust may produce an important signature.  In the standard birth ring model, dust will eventually assume eccentric orbits as the radiation pressure unsuccessfully attempts to blow out grains just above the blow-out radius.  Smaller grains will assume a larger eccentricity, smearing the total distribution over a larger radial range.  Dust at different radii in the disc will assume a different equilibrium temperature with the stellar radiation, affecting the overall emission from the system.  TAM missions will preferentially select asteroids closer to the planet of origin, cf Near Earth Objects (NEOs) as opposed to the Main Belt.  Dust production will therefore initially occur in a different location (although this may still correspond closely with the birth ring).  Depending on the debris produced by mining, an anomalous dust source will appear in the system, which may have a (small) measurable effect on the SED produced from the debris disc (or detectable through high-resolution imaging).  Equivalently, civilisations may consider collecting exozodiacal dust as a source of ground silicates, which would produce an anomalous dust sink (although again the large quantities of silicates in terrestrial planet crusts would appear to make this unnecessary).

Drilling into asteroids on a large scale requires large inputs of energy.  Any debris thrown off during drilling may be strongly heated and reformed.  Glassy silicas such as obsidian and tektites may be detected, as well as fluorescing silicon oxide (SiO), a byproduct from the vapourisation of silicates. However, these have also been modelled in hypervelocity systems \citep{Lisse2009}.  Dust produced during mining will also have an initial thermal energy imparted to them.  If this thermal energy is high enough, this may also have an effect on SEDs and imaging, requiring models to assume an unusual temperature gradient in the disc.   

Heated dust will cool on short timescales (of order a few minutes).   If mining is prolific in the system, then there may be variability in the system's flux at a given wavelength, with periodic fluctuations correlated with the cooling time.  If detected, the strength of these fluctuations could be used to model the artificial dust production rate, and ultimately estimate the local mining rate in the system.   
However, these fluctuations will be extremely small compared to the mean signal, and will require very high-cadence observations to detect them.  

Finally, mining strategies may require operations to begin and cease for various reasons (either technical or political).  If several asteroids are being mined simultaneously (where the start and stop of mining are uncorrelated), then this will introduce anomalous variability at differing locations in the disc, which may be resolved by imaging.  If the starts and stops are highly correlated (which may be the case if political motivations are important), then this variability may be strong enough to be detectable by SED.  This would require multi-epoch observations of a given target, which will probably not be feasible with current instrumentation and user demands.  These signatures will presumably have to wait for future instrumentation if they are to be detectable.

\section{Conclusions}

\noindent This paper has studied the hypothesis that targeted asteroid mining (TAM) may leave observational signatures in debris discs, and be detectable by astronomers observing them in the infrared and other wavelengths.  We argue that TAM may be a common milestone in the development of space-faring civilisations, and therefore if intelligent civilisations are common, then these observational signatures would also be common.  We have considered three classes of disequilibrium induced by TAM in debris discs.  Firstly, chemical disequilibrium induced by the extraction of specific minerals and elements, secondly the mechanical disequilibrium induced by deliberately destroying and mining the larger asteroids in the system, and lastly thermal disequilibrium induced by producing dust with unusual temperature distributions as a result of mining.

The general trend is somewhat disappointing.  For TAM to be detectable, it must be prolific and industrial-scale, producing a large amount of debris and disrupting the system significantly to be detected.  However, instrumentation is continually improving, and sensitivity to such effects will only grow, reducing the constraints on detectability.  What remains indefatigable with technological advance is the confusion of apparent TAM signals with natural phenomena.  A detection of any one of these TAM signals can be explained with a simpler natural model, but detection of many (or all) of these signals in tandem will prove more difficult to model, and hence TAM more difficult to discount as a possibility.

We cannot therefore expect a conclusive detection of extraterrestrial intelligence (ETI) by TAM - what it can provide is a call to attention.  Debris disc systems with unusual dust size distributions and locations, or deficits in chemical composition provide astrobiologists with candidates for further study.  Further characterisation of the planets in the debris disc system and an assessment of their habitability will ultimately be a better measure of the likelihood of ETI, but tentative signals of TAM may be the first clue that alerts us to the possibility of their existence.  Indeed, TAM signals would be a stronger indication of ETI than say, biomarkers on an exoplanet (e.g. \citealt{Kaltenegger2010}).  While biomarkers are a strong method of identifying life-forms in general, TAM requires intelligent, technological life-forms.  For these reasons, searching for TAM through ``piggy-back'' methods should be considered as another string to the bow of a multi-wavelength, multi-signal SETI enterprise.

\bibliographystyle{mn2e} 
\bibliography{ija_duncanforgan}

\end{document}